\documentclass[showpacs,twocolumn]{revtex4}
\usepackage[dvips]{epsfig}

\newcommand{\be}{\begin{equation}}
\newcommand{\ee}{\end{equation}}
\newcommand{\clF}{{\cal F}}

\newcommand{\clL}{{\cal L}}

\newcommand{\clE}{\mathcal{E}}

\newcommand{\clM}{{\cal M}}

\newcommand{\bea}{\begin{eqnarray}}
\newcommand{\eea}{\end{eqnarray}}

\newcommand{\prt}{\partial}

\newcommand{\tlnu}{\tilde{\nu}}
\newcommand{\rgl}{\rangle}
\newcommand{\lgl}{\langle}

\begin{document}

\title{Superdiffusive comb: Application to experimental
observation of anomalous diffusion in one dimension}
\author{Alexander Iomin}

\affiliation{Department of Physics, Technion, Haifa, 32000,
Israel}

\date{PHYSICAL REVIEW E 86, 032101 (2012)}

\begin{abstract}
A possible mechanism of superdiffusion of ultra-cold atoms in a
one-dimensional polarization optical lattice, observed
experimentally in [Phys. Rev. Lett. \textbf{108}, 093002 (2012)],
is suggested. The analysis is based on a consideration of
anomalous diffusion in a fractal comb [Phys. Rev. E \textbf{83},
052106 (2011)]. It is shown that the transport exponent is
determined by the fractal geometry of the comb due to recoil
distributions resulting in L\'evy flights of atoms.

\end{abstract}


\maketitle

Recently, $1D$ heavy-tailed distributions were observed in
experimental studies of anomalous diffusion of ultra-cold
${}^{87}{\rm Rb}$ atoms in a one-dimensional optical lattice
\cite{sagi2011}. It was found that the initial ensemble of atoms
spread superdiffusively, such that the ``full widths at half the
maximum'' (FWHM) increases with time like $t^{\frac{1}{\mu}}$ with
diffusion exponent $1<\mu<2$. Another important observation was
the dependence of the transport exponent on the depth of the
lattice potential \cite{sagi2011}. The theoretical explanation of
this fact, presented within the standard semiclassical treatment
of Sisyphus cooling \cite{kb2012}, is based on a study of the
microscopic characteristics of the atomic motion in optical
lattices and recoil distributions resulting in macroscopic L\'evy
flights in space, such that the L\'evy distribution of the flights
$g(l)$ depends on the lattice potential depth \cite{mez1996}:
$g(l)\sim l^{-1-\nu}$ with $\nu=\frac{1+D}{3D}$, where
$D=cE_R/U_{\rm trap}$, while $U_{\rm trap}$ is the lattice
potential depth scaled by recoil energy $E_R$, and $c$ is a
dimensionless parameter (adopting notation from
Ref.~\cite{kb2012}). A relation between the diffusion exponent
$\mu$ and the L\'evy distribution exponent $\nu$ was established
for different regimes of the atomic dynamics, which is described
by Fokker- Planck dynamics in an asymptotically logarithmic
potential \cite{kb2012,dlkb2011}.

\begin{figure}
\begin{center}
\epsfxsize=9.0cm \leavevmode
    \epsffile{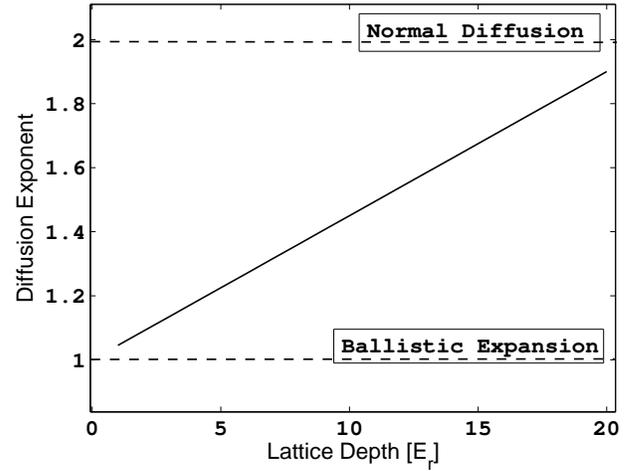}
\caption{A sketch of experimental observation  of a wave packet
spreading in a 1D optical lattice \cite{sagi2011}. The diffusion
(fractional) exponent, presented vs the lattice potential depth,
describes the full width at half the maximum (FWHM) of the atom
distribution: FWHM $\,\sim t^{\frac{1}{\mu}}$. The dashed lines
correspond to the limiting cases.}
\end{center}
\end{figure}

Here, a relation between the transport exponent and the lattice
potential depth is established in the framework of a fractional
comb model \cite{iom2011}. We analyze an experimental observation
of dependence of a diffusion exponent as a function of the lattice
depth $U_{\rm trap}$, presented in Fig.~3 of Ref.~\cite{sagi2011}
(see here Fig.~1), where the limiting cases correspond to normal
diffusion and ballistic motion, and these can be easily modelled
by turbulent diffusion on a comb. This anomalous diffusion is
described by the $2D$ distribution function $P=P(x,y,t)$, and a
special behavior is that the displacement in the $x$--direction is
possible only along the structure axis ($x$-axis at $y=0$), while
any spreading along the $x$ direction inside the fingers (motion
along the $y$ direction) is not possible. The Fokker-Planck
equation in some dimensionless variables reads ( see
\textit{e.g.}, \cite{iom2011})
\be\label{lc_1}  %
\prt_tP= \delta(y) \prt_{|x|}P +d\prt_y^2P\, , \ee  %
where $\prt_{|x|}$ is a particular case of the Riesz derivative,
which can be defined by its Fourier image (see \textit{e.g.},
\cite{zaslavsky})
\be\label{lc_1a} %
\hat{\clF}[\prt_{|x|}^qf(x)]=-|k|^q\hat{f}(x)\ee  %
and $\hat{\clF}[f(x)]=\hat{f}(k)$ with $q=1$ in Eq. (\ref{lc_1}).
It should be admitted that the $y$ axis is the auxiliary space,
introduced for the trap modelling. In other words, it is
introduced to model a non-Markovian process by means of Markovian
description. The true distribution is the distribution function
along the $x$ axis, which is
\be\label{lc_1b} %
\overline{P}(x,t)=\int_{-\infty}^{\infty}P(x,y,t)dy\, .\ee %

To establish the connection between turbulent diffusion on the
comb with anomalous diffusion of cold atoms, we, first, model the
limiting cases, shown in Fig.1. An effective constant diffusion
coefficient in the $y$ direction can be considered as a function
of the lattice depth $d=d(U_{trap})$. Therefore, Eq. (\ref{lc_1})
describes the two limiting cases (dashed lines in Fig.~1), where
we use that $d=\Theta(U_{\rm trap})$ is the Heaviside function.
One easily checks \cite{bi2004} the full width at half the maximum
(FWHM) of the atom distribution. For $U_{\rm trap}=0$ one has FWHM
$=\sqrt{\lgl x^2\rgl}=t$ that corresponds to ballistic motion,
while for $d=1$ one finds normal diffusion with FWHM
$=t^{\frac{1}{2}}$. Here we use the definition
\be\label{lc_2} %
\lgl x^2(t)\rgl=\int_{-\infty}^{\infty}x^2\overline{P}(x,t)dx\, . \ee %

For the ``shallow'' trap lattice potentials, we modify Eq.
(\ref{lc_1}) in the following way. From the experimental
realization we know that after photon emission due to recoil atoms
``fly'' on distances distributed by power law and, correspondingly
the fingers, as the traps, are distributed by power law with the
fractal dimension $\tlnu$, related to the L\'evy flights.
Therefore, this experimental realization can be described by the
fractal comb model, developed in \cite{iom2011}. Following this
consideration, one can consider this set of the traps as a fractal
set $F_{\tlnu}(x)$ with the fractal dimension $\tlnu$, which is
embedded in the $1D$ space, $0<\tlnu<1$. Therefore, the effective
diffusion coefficient becomes inhomogeneous $d\rightarrow
d\chi(x)$, where $\chi(x)$ is a characteristic function of
$F_{\tlnu}(x)$, such that $\chi(x)=1$ for $x\in F_{\tlnu}(x)$ and
$\chi(x)=0$ for $x\notin F_{\tlnu}(x)$. Taking this into account,
we modify Eq. (\ref{lc_1}) in the form
\be\label{lc_0b}  %
\prt_tP= \delta(y) \prt_x^2P +d\chi(x)\prt_y^2P\, . \ee  %
To arrive at the corresponding modification of Eq (\ref{lc_1}), we
apply the Fourier transform to Eq. (\ref{lc_0b}) with respect to
the $x$ coordinate. To this end we use the auxiliary identity
\be\label{lc_0c}  %
\chi(x)f(x)\equiv\prt_x\int_{-\infty}^x\chi(y)f(y)dy\equiv
-\prt_x\int_x^{\infty}\chi(y)f(y)dy\ee %
with the boundary conditions $P(x=\pm\infty)=0$. This integration
with the characteristic function can be carried out by means of a
convolution \cite{bi2011a,iom2011,Ren}
\be\label{lc_0d}   %
\int_{-\infty}^x\chi(y)f(y)dy\Rightarrow
{}_{-\infty}I_x^{\tlnu}f(x)=
\int_{-\infty}^x\frac{f(y)(x-y)^{\tlnu-1}dy}{\Gamma(\tlnu)}\, , \ee %
where $\Gamma(\tlnu)$ is the Gamma function and we also use the
convenient notations of fractional integration:
${}_{-\infty}I_x^{\tlnu}f(x)$
\cite{zaslavsky,klafter,SokKlafBlum}. We also used here the
following arguments for the characteristic function. Note that
$$
\int_{-\infty}^{\infty}\chi(y)f(y)dy=\sum_{x_j\in
F_{\alpha}}\int_{-\infty}^{\infty}f(y)\delta(y-x_j)dy\, , $$ where
$$\sum_{x_j\in F_{\alpha}}\delta(y-x_j)=\clM^{\prime}(x)\sim
|x|^{\alpha-1}$$ is a local fractal density, such that
$\int_{-x}^xd\clM(y)\sim |x|^{\alpha}$ corresponds to the fractal
volume. Therefore, due to Theorem $3.1$ in Ref.~\cite{Ren} we have
$\int_0^xf(y)d\clM(y)\simeq
\frac{1}{\Gamma(\alpha)}\int_0^x(x-y)^{\alpha-1}f(y)dy$.

Using this fractional integration one obtains from Eq.
(\ref{lc_0c}) the fractional derivative
$\prt_x[{}_{-\infty}I_x^{\tlnu}P(x,y,t)]$ defined in the
Riemann-Liouville form \cite{zaslavsky,klafter,SokKlafBlum}.
Performing the Fourier transform and taking the symmetrical form,
one obtains the following change in Eq. (\ref{lc_1}) $d\rightarrow
d|k|^{1-\tlnu}$. Note that $\tlnu=\tlnu(U_{\rm trap})$ is the
fractal dimension of the fingers distribution on the $x$ axis.
After this change Eq. (\ref{lc_1}) reads
\be\label{lc_3} %
\prt_t\hat{P}=-\delta(y)|k|\hat{P}+d|k|^{2-\nu}\prt_y^2\hat{P}\, , \ee %
where $\nu=1+\tlnu$ corresponds to the L\'evy distribution related
to the lattice potential depth \cite{mez1996}. To satisfy the
limiting cases, we have $\nu=2$ and $d=\Theta(U_{\rm trap})$ is
the Heaviside function.

Our aim now is to find $\overline{P}(x,t)$, defined in Eq.
(\ref{lc_1b}) for ``shallow'' trap lattice potentials, when $
1<\nu<2$ ($\frac{1}{5}<D<\frac{1}{2}$). To this end, an analysis
developed in \cite{iom2011} is applied. One carries out the
Laplace transform in the time domain
$\hat{\clL}[\hat{P}(k,y,t)]=\hat{\tilde{P}}(k,y,s)$. Looking for
the solution of the Laplace image in the form
\be\label{lc_4} %
\hat{\tilde{P}}(k,y,s)=\exp[-|y|\sqrt{|k|^{\nu-2}s/d}]f(k,s)\,
,\ee  %
one arrives at the intermediate expression in the form of the
Laplace and Fourier inversions
\be\label{lc_5} %
 P(x,y,t)=\hat{\clF}_k^{-1}\left\{\hat{\clL}^{-1} \left[
\frac{2e^{-|y|\sqrt{s|k|^{\nu-2}/d}}}{2\sqrt{sd |k|^{2-\nu}}+
|k|}\right]\right\}\, .\ee %
Integration over $y$ and the inverse Laplace transform yield a
solution in the Fourier inversion form:
\be\label{lc_6} %
\overline{P}(x,t) =\frac{1}{2\pi}\int_{-\infty}^{\infty}e^{-ikx}
\clE_{\frac{1}{2}}\Big(-\frac{1}{2}\sqrt{|k|^{\nu}t/d}\Big)dk\,
.   \ee %
Here $$\clE_{\alpha}(-z)=\frac{1}{2\pi
i}\int_{\gamma}\frac{u^{\alpha-1}e^udu}{u^{\alpha}+z}$$ is the
Mittag-Leffler function defined by the inverse Laplace transform
with a corresponding deformation of the contour of the integration
\cite{batmen}. First, we admit the scaling variable
${x}/{t^{\nu}}$ that corresponds to the superdiffusion expansion
$$
{\rm FWHM}\sim t^{\frac{1}{\nu}}\, .$$  This also corresponds to
the experimental observations, presented in Ref.~\cite{sagi2011},
and to the scaling obtained in Ref.~\cite{kb2012} (see Eq. (11)
there).

Let us consider an asymptotic behavior of this superdiffusion
expansion of the initial packet of ultra-cold atoms at
$x\rightarrow\infty\,(k\rightarrow 0)$. In this case, the argument
of the Mittag-Leffler function is small, yielding \cite{klafter}
$$\clE_{\alpha}(z)\sim\exp\Big(-\frac{z}{\Gamma(1+\alpha)}\Big)\,
,$$ and one obtains
$$ \overline{P}(x,t)
=\frac{1}{2\pi}\int_{-\infty}^{\infty}e^{-ikx}
\exp\Big(-\frac{\sqrt{|k|^{\nu}t}}{3\sqrt{d\pi}}\Big)dk\, .$$ %
This is nothing but the Fourier inversion of the the
characteristic function of a centered and symmetric L\'evy
distribution \cite{klafter,bouchaud} that describes L\'evy flights
of atoms. This integration yields the analytical solution in the
form of the Fox function with the power-law asymptotics
\cite{klafter,bouchaud}
\be\label{lc_9}  %
\overline{P}(x,t)\sim \frac{t}{|x|^{1+\nu}}\, , ~~ \nu<2\, .
\ee  %
From here one obtains for the first moment \cite{klafter,bouchaud}
$\lgl x(t)\rgl\sim t^{\frac{1}{\nu}}$, which corresponds to the
scaling obtained above.

In conclusion, we note that this description is a particular case
of a general scheme of the wave-packet spreading described by the
fractional Fokker-Planck equation (FFPE) that can be obtained for
the true distribution (\ref{lc_9}). Integrating the Laplace image
of Eq. (\ref{lc_0b}) over $y$, and taking into account Eq.
(\ref{lc_4}), one performs the Laplace and the Fourier inversions
that yields the FFPE for the true distribution
\be\label{lc_10} %
\prt_t^{\frac{1}{2}}\overline{P}(x,t)=
\frac{1}{2\sqrt{d}}\prt_{|x|}^{\frac{\nu}{2}}\overline{P}(x,t)\, . \ee  %
Here $\prt_t^{\frac{1}{2}}$ is the Caputo time fractional
derivative \cite{mainardi} and $\prt_{|x|}^{\frac{\nu}{2}}$ is the
Riesz fractional derivative \cite{zaslavsky} defined in Eq.
(\ref{lc_1a}). This equation is a particular case of the FFPE
\be\label{lc_11} %
\prt_t^{\alpha}\overline{P}=K(\alpha,d)\prt_{|x|}^q\overline{P}\,
,
\ee %
where $K(\alpha,d)$ is a generalized diffusion coefficient and the
scaling $q/\alpha=\nu$ is fulfilled for the initial ensemble of
atoms, which spreads superdiffusively with $\alpha<1$ and $q<2$.

\end{document}